\documentclass{nature}
\usepackage{graphicx,amsmath}
\graphicspath{{}{./}}
\DeclareGraphicsExtensions{.pdf,.png,.jpg}

\usepackage{graphicx}

\newenvironment{narrow}{\baselineskip=7mm}

\newenvironment{narroww}{\baselineskip=5mm}

\def\ga{\ \lower 3pt\hbox{${\buildrel > \over \sim}$}\ }
\def\la{\ \lower 3pt\hbox{${\buildrel < \over \sim}$}\ }
\newcommand{\st}{\rm St}

\title{Uranian Satellite Formation by Evolution of a Water Vapor Disk Generated by a Giant Impact}

\author{Shigeru Ida,$^{1\ast}$ Shoji Ueta,$^{2}$ Takanori Sasaki,$^{3}$ and Yuya Ishizawa$^{3}$}

\begin{document}

\maketitle

\begin{affiliations}
\begin{small}
\begin{narrow}
 \item ELSI, Tokyo Institute of Technology, Ookayama 2-12-1, Meguro-ku, Tokyo 152-8550, Japan, $^\ast$ E-mail: ida@elsi.jp
\item Graduate School of Advanced Integrated Studies in Human Survivability, Kyoto University, Nakaadachi-cho 1, Yoshida, Sakyo-ku, Kyoto 606-8306, Japan
\item Department of Astronomy, Kyoto University, Kitashirakawa-Oiwake-cho, Sakyo-ku, Kyoto 606-8502, Japan
\end{narrow}
\end{small}
\end{affiliations}

\begin{narrow}

\begin{abstract}
The ice-giant planet Uranus likely underwent a giant impact, given that its spin axis is tilted by 98 degrees 
\cite{Slattery92, Kurosaki19,Reinhardt19}.
That its satellite system is equally inclined and prograde suggests that it was formed as a consequence of the impact. However, the disks predicted by the impact simulations\cite{Slattery92, Kegerreis18,Reinhardt19} generally have sizes one order smaller and masses two orders larger than those of the observed system at present. Here we show, by means of a theoretical model, that the Uranian satellite formation is regulated by the evolution of the impact-generated disk. Because the vaporization temperature of water ice is low and both Uranus and the impactor are assumed to be ice-dominated, we can conclude that the impact-generated disk has mostly vaporized. We predict that the disk lost a significant amount of water vapour mass and spread to the levels of the current system until the disk cooled down enough for ice condensation and accretion of icy particles to begin. From the predicted distribution of condensed ices, our N-body simulation is able to reproduce the observed mass-orbit configuration of Uranian satellites. This scenario contrasts with the giant-impact model for the Earth's Moon\cite{Canup01}, in which about half of the compact, impact-generated, solid or liquid disk is immediately incorporated into the Moon on impact\cite{Ida97}.
\end{abstract}

Uranus has five major satellites in a mass range of $10^{-6}$--$10^{-4} \, M_{\rm U}$ (Fig.~1), where $M_{\rm U} \simeq 8.7 \times 10^{25}\, {\rm kg}$ is Uranus mass, extended to $\sim 25 \, r_{\rm U}$,
where $r_{\rm U} \simeq 2.5\times 10^7\,{\rm m}$ is Uranus' physical radius (Fig. 1).
The extension to $\sim 25 \, r_{\rm U}$ cannot be accounted for by tidal orbital expansions \cite{Dermott88}.
Their orbits are prograde to Uranus' spin and nearly circular.
The total mass of the satellites is $\sim 10^{-4} \, M_{\rm U}$.
The rock to ice ratios of the satellites are observationally estimated to be nearly about 1:1 except 
the innermost Miranda \cite{Hussmann06}, while Uranus consists mostly of ices \cite{Podolak95}
For formation of the satellites,
the impact \cite{Slattery92} and the circum-planetary sub-disk scenarios have been proposed \cite{Szulagyi18}.
Because the sub-disk that feeds
H/He gas from a circum-stellar disk to the planet would be formed on
the planetary orbital plane, the sub-disk scenario does not reconcile with
the inclined satellite system, 
unless multi-step complicated mechanisms are considered
\cite{Morbidelli12}.
It is simple to consider that the satellites
are formed in the disk generated by the impact
that tilted the spin axis and caused the current spin period ($\sim 17.2$ hours).
The accretion of the satellites from the impact-generated disk naturally 
results in the prograde orbits on Uranus' equatorial plane.
However, the theoretically predicted 
impact-generated disks 
\cite{Slattery92, Kegerreis18,Reinhardt19} 
are not only one order smaller and two orders more massive than the current system
but also are significantly depleted in rocky components, because rocks in the small core are not easily ejected by the impact.
These difficulties were raised by
the past papers \cite{Slattery92, Kegerreis18, Reinhardt19} 
by the simple translation of the giant impact model of the Earth's Moon \cite{Canup01} to Uranus
with the lack of following of the evolution of the water vapor disk.

\begin{figure}
\includegraphics[width=140mm]{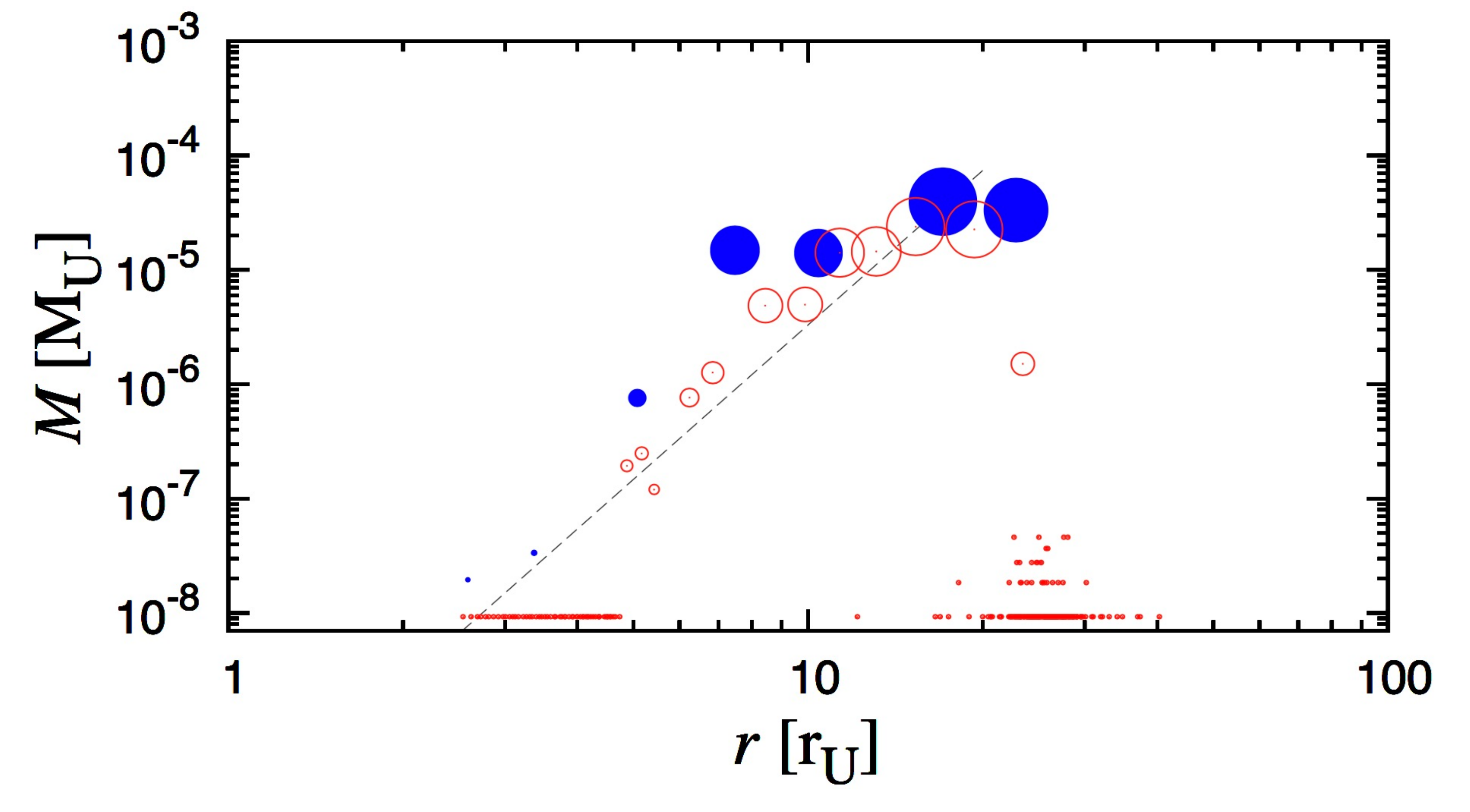}
\caption{\narroww
The mass ($M$) and orbital radius ($r$) distribution of 
the current Uranian satellite systems and that predicted by N-body simulation. 
The five major Uranian satellites are represented by
the filled blue circles in the range of $\ga 10^{-6} M_{\rm U}$,
where $M_{\rm U}$ and $r_{\rm U}$ are the mass and orbital radius of Uranus. 
Minor satellites with $10^{-8}-10^{-7} M_{\rm U}$ are also plotted
with tiny filled blue circles.
The size of the circles is proportional to the physical radius.
The open red circles represent the result of N-body simulations
of accretion from condensed icy particles 
(10000 bodies with masses of $0.92 \times 10^{-8} M_{\rm U}$)
at 1300 years (see Methods).
With a longer run, some of the accreted satellites would collide with each other,
minor satellites would accrete from the small satellitesimals 
with $M  \sim 10^{-8} M_{\rm U}$ at $r < 10 \, r_{\rm U}$
and the satellitesimals with $M  \sim 10^{-7} M_{\rm U}$ at $r > 10 \, r_{\rm U}$ would be swept by the proto-satellites, which is more consistent with the current Uranian satellites.  
The dashed black line is analytically derived ``isolation mass"
in oligarchic growth model \cite{Kokubo00} given by Eq.~(\ref{eq:miso0}).
}
\label{fig:Msat}
\end{figure}

We assume that both Uranus and the impactor are
ice-dominated with small rocky cores 
and that Uranus is covered by 3--10 wt.\% of H/He atmosphere.
The Uranus gravity accelerates the impact velocity to $\ga 20 \,{\rm km/s}$, equivalently, the impact energy to $\ga 2 \times 10^8 {\rm J/kg}$, which
is 100 times larger than the latent heat of H$_2$O ice.
As a result, the impact-generated disk consists of a mixture of 
water vapor and H/He gas.
Although the icy mantle also includes CH$_4$ and NH$_3$ ices,
we only consider the most abundant ice, H$_2$O, as a representative of the ices.
Since $(c_s/v_{\rm K})^2 \simeq 3.3 \times 10^{-2} (\mu_{\rm all}/2.8)^{-1}
(T/10^4 \, {\rm K}) (r/r_{\rm U}) \ll 1$, 
where $T$ is the disk temperature, $\mu_{\rm all}$ is the mean molecular weight of the mixture, $c_s$ is the local sound velocity, $v_{\rm K}$ is the local Keplerian velocity,
the evaporated vapor does not escape from the Uranian system
and stays as a circum-planetary disk.  
% As we show below, most of the vapor is accreted to the planet and the small amount of residual vapor condenses at $\sim 20 \, r_{\rm U}$.

As we will show below, 
the final satellite mass and orbital distributions are solely determined by
a condensation sequence of icy grains in the disk,
% of a mixture of water vapor and H/He gas
and the turbulent viscous spreading and cooling of the disk play an essential role in the satellite formation.
We numerically solve 
the 1D viscous diffusion equation of disk gas surface density $\Sigma_{\rm g}$, given by \cite{Hartmann98}
\begin{equation}
\frac{\partial \Sigma_{\rm g}}{\partial t} - 
{1 \over r} \frac{\partial}{\partial r}
\left[ 3 r^{1/2} \frac{\partial}{\partial r} (\Sigma_{\rm g} \nu r^{1/2}) 
\right] = 0,
\label{eq:surf_density_evol2}
\end{equation}
where the turbulent kinetic viscosity 
is modeled by $\nu = \alpha c_s^2 \Omega^{-1}$,
where $c_s$ and $\Omega$ are the local sound velocity and 
orbital frequency of the disk gas, and $\alpha$ is
a constant parameter to represent the turbulence strength ($\alpha \ll 1$) \cite{Shakura73}.  
As local disk temperature, we use the photo-surface temperature by the viscous heating for simplicity \cite{Hartmann98}, 
\begin{equation}
T \simeq \left(\frac{9 \, GM_{\rm U} \Sigma_{\rm g} \nu}{8 \,\sigma r^3} \right)^{1/4},
\label{eq:Tvis}
\end{equation}
where $G$ is the gravitational constant and $\sigma$ is the Stefan-Boltzmann constant. 

\begin{figure}
\begin{center}
\includegraphics[width=180mm]{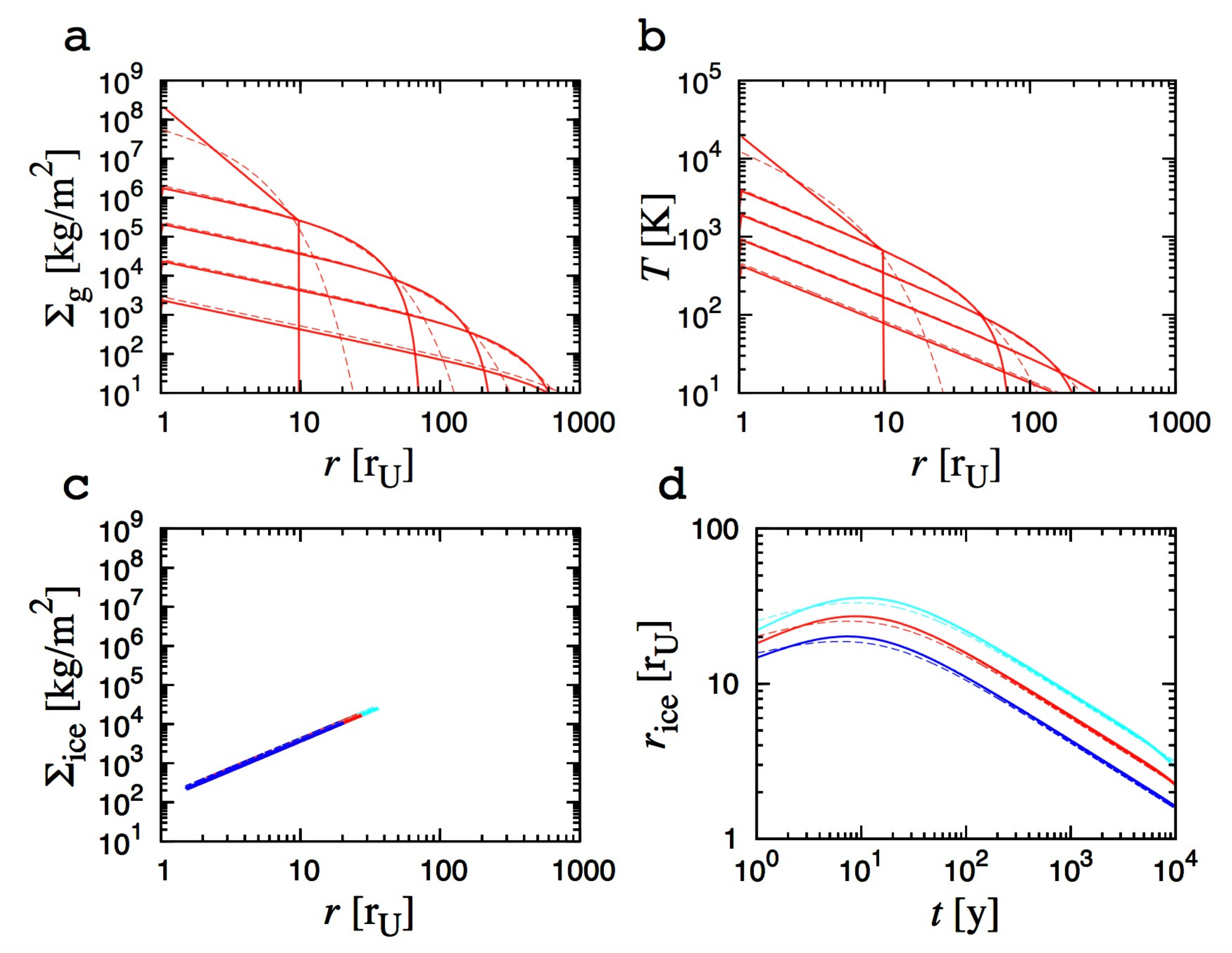}
\end{center}
\caption{\narroww
The evolution of the disk of a mixture of H/He gas and
water vapor and the associated ice condensation:
(a) The disk gas surface density ($\Sigma_{\rm g}$);
(b) the disk temperature ($T$) with $\alpha = 10^{-3}$.
The solid and dashed red lines are
the numerically solved distribution and analytical one
(Eqs.~(\ref{eq:Tvis}) and (\ref{eq:Sigma_selfsim0})),
In panels a and b, the upper to lower curves for $r < 10\, r_{\rm U}$ 
represent
the distributions at $t = 0, 10, 10^2, 10^3$ and $10^4$ years.
The initial disk for the numerical calculation is set as a centrally-confined one,
$\Sigma_{\rm g,imp} = 2.4 \times 10^8 (r/r_{\rm U})^{-3} \, \rm kg/m^2$ with a truncation at $r=10\, r_{\rm U}$, which has $M_{\rm d,imp} = 10^{-2} M_{\rm U}$ and $\langle r_{\rm d,imp} \rangle \simeq 2.3\,  r_{\rm U}$. 
In the analytical self-similar formula, $r_{\rm d0}=3 \, r_{\rm U}$ and 
$\Sigma_{\rm d0} = 0.3 \, \Sigma_{\rm g,imp}$ are used, according to
the conversion given by 
Eqs.~(\ref{eq:r_g0}) and (\ref{eq:Sig_g0}).
The time evolution of the ice line is plotted in panel d.
The blue, red, and right blues lines are for 
$M_{\rm d,imp} = 3\times 10^{-3} M_{\rm U},
10^{-2} M_{\rm U}$, and $3\times 10^{-2} M_{\rm U}$, respectively.
When $T$ becomes equal to $T_{\rm ice}$, we assume that ice condenses with 
the surface density $\Sigma_{\rm ice}=\gamma \Sigma_{\rm g}$ at that time
(panel c), where we assumed $\gamma = 0.3$.
 }
\end{figure}

The numerically solved $\Sigma_{\rm g}$ and $T$ evolution of the disk
is shown in Figs.~2a and b.
They show that the impact-generated disk quickly spreads and cools.
By the total angular momentum conservation, the spreading is associated
by accretion of the disk onto the planet. 
The disk converges to a quasi-steady accretion disk
where the $\Sigma_{\rm g}$ and $T$ distributions with the radial distance $r$ evolve self-similarly.
We derive an approximate expression for the self-similar solution of $\Sigma_{\rm g}$ and $T$
in order to generalize the numerical results.
For steady accretion ($\Sigma_{\rm g} \nu$: const.), $T \propto r^{-3/4}$ (Eq.~(\ref{eq:Tvis})) and
$\nu \propto c_s^2 \Omega^{-1} \propto T r^{3/2} \propto r^{3/4}$.
The self-similar solution to the above equation with time independent
$\nu$ was already derived 
\cite{Lynden-Bell74, Hartmann98}.
In our case, 
$\nu$ also depends on $\Sigma_{\rm g}$ through $T$ (Eq.~(\ref{eq:Tvis}))
and it decreases with time.
We modify the original self-similar solution incorporating  
the additional $\Sigma_{\rm g}$-dependence as (Methods) 
\begin{eqnarray}
\Sigma_{\rm g} & = & \Sigma_{\rm g,U0} \,t_{*0}^{-21/22} \, 
\left(\frac{r}{r_{\rm U}}\right)^{\,-3/4}
\exp \left[ - \left(\frac{r}{r_{\rm d0}\,t_{*0}^{-12/22}} \right)^{5/4} \right], 
\label{eq:Sigma_selfsim0} \\
t_{*0} & = & 1+\frac{t}{t_{\rm diff0}} = 1 + \frac{t}{(16/75)
( r^2/\nu)_{r_{\rm d0},t=0}},
\end{eqnarray}
where $\Sigma_{\rm g,U0}$ is the disk gas surface density at $r=r_{\rm U}$
and at $t=0$,  
$t_{\rm diff0}$ is the viscous diffusion timescale at $r_{\rm d0}$,
and $r_{\rm d0}$ is the characteristic disk radius at $t=0$, respectively.
We define $t$ as the time after the impact-generated disk is relaxed to 
the quasi-steady self-similar solution.
The corresponding analytical $T$ is derived
from $\Sigma_{\rm g}$ with Eq.~(\ref{eq:Tvis}).
The analytical solutions reproduce the numerical results
except for the parts of the exponential tail (Figs.~2a and b).

The values of $r_{\rm d0}$ and $\Sigma_{\rm g,U0}$
in the analytical solutions 
are given by the quantities of the impact-generated disk as
(Eqs.~(\ref{eq:rd0_jdisk}) and (\ref{eq:Sigmag0_jdisk}) in Methods)
\begin{eqnarray} 
r_{\rm d0} & \simeq & 
3.0 \left(\frac{\langle r_{\rm d,imp} \rangle}{2\, r_{\rm U}}\right) \,r_{\rm U}, \label{eq:r_g0}\\ 
\Sigma_{\rm g,U0} & 
\simeq & 6.5 \times 10^7 \left(\frac{\langle r_{\rm d,imp} \rangle}{2\, r_{\rm U}}\right)^{-5/4}
                              \left(\frac{M_{\rm d,imp}}{10^{-2} M_{\rm U}}\right) \,\rm kg/m^2,
\label{eq:Sig_g0} 
\end{eqnarray}
where $M_{\rm d,imp}$ is the total mass of the impact generated-disk,  
$\langle r_{\rm d,imp}\rangle$ is its mean orbital radius defined by 
$\langle r_{\rm d,imp} \rangle = ((J_{\rm d,imp}/M_{\rm d,imp})/r_{\rm U}^2 \Omega_{\rm U})^2 \, r_{\rm U}$, $J_{\rm d,imp}$ is its total 
angular momentum, and 
$\Omega_{\rm U}$ is the disk orbital frequency at $r=r_{\rm U}$.
Thus, it is demonstrated that
the disk spreading and cooling are mostly determined
by only two parameters, $\langle r_{\rm d,imp} \rangle$
and $M_{\rm d,imp}$, independent of other details of the impact-generated disk.
The past impact simulations \cite{Slattery92, Kegerreis18,Reinhardt19}
showed that $\langle r_{\rm d,imp} \rangle \sim 2\, r_{\rm U}$ and
$M_{\rm d,imp} \sim 10^{-2} M_{\rm U}$ are typical values.

When the disk temperature decays to the ice condensation
temperature $T_{\rm ice} \simeq 240\,{\rm K}$ (Eq.~(\ref{eq:T_snow}) in Methods)
for the first time, we deposit the condensed ice surface density
by $\Sigma_{\rm ice}=\gamma \, \Sigma_{\rm g}$, where $\gamma$ is 
the abundance of water vapor in the disk.
SPH (Smooth Particle Hydrodynamics) simulations suggest $\gamma \sim 0.1-0.5$
\cite{Slattery92, Kegerreis18,Reinhardt19}.
We use $\gamma = 0.3$ as a nominal value and $\gamma_{03} = \gamma/0.3$.
With $T \simeq 240\,{\rm K}$, the numerically obtained
$\Sigma_{\rm ice}$ and deposited radius (``ice line")  $r_{\rm ice}$ 
are plotted in Figs.~2c and d.
Because ice condensation occurs after significant evolution of the quasi-steady disk, the ice distribution is independent of detailed structure of the initial impact-generated disk.
In particular, $\Sigma_{\rm ice}$ at each $r$ is independent of $M_{\rm d,imp}$
(Fig.~2c), and the analytical estimation of $\Sigma_{\rm ice}$ below
shows that it is independent even of $\langle r_{\rm d,imp} \rangle$.
From Eq.~(\ref{eq:Tvis}),
\begin{equation}
T \simeq 240 \left(\frac{\alpha}{10^{-3}}\right)^{1/3} \left(\frac{\Sigma_{\rm g}}{4.0 \times 10^2 \, {\rm kg/m^2}}\right)^{1/3}
\left(\frac{r}{r_{\rm U}}\right)^{-1/2} 
\; \rm K. 
\label{eq:T_Sigma}
\end{equation}
From Eq.~(\ref{eq:T_Sigma}) with $T_{\rm ice} \sim 240\,{\rm K}$, we obtain
\begin{equation}
\Sigma_{\rm ice} \simeq \gamma \Sigma_{\rm g} 
\simeq 1.2 \times 10^{2} \beta^{-1}\gamma_{03}
\left(\frac{r}{r_{\rm U}}\right)^{3/2} \; {\rm kg/m^2}, \label{eq:Sigma_ice}
\end{equation}
where $\beta  = (\alpha/10^{-3}) (T_{\rm ice}/240\,{\rm K})^{-3}$.
This completely reproduces 
$\Sigma_{\rm ice}$ by the numerical solution (Fig.~2c).

The positive gradient of $\Sigma_{\rm ice} \, (\propto r^{3/2})$ 
is produced from $\Sigma_{\rm g}$ with the negative slope $(\propto r^{-3/4})$,
because, in inner regions, the viscous heating is more efficient 
(Eq.~(\ref{eq:Tvis})) and the disk must be more significantly depleted
to realize $T \la T_{\rm ice}$ than in outer regions.
The positive gradient implies that
most of the condensed ice mass is located in an outermost region.
While $\Sigma_{\rm ice}$ does not depend on $\langle r_{\rm d,imp} \rangle$
and $M_{\rm d,imp}$ at each $r$,
they affect how far the distribution extends, although the dependences are weak.
The outer truncation radius for the $\Sigma_{\rm ice}$-distribution is 
evaluated as below and it reproduces the numerical results.

The ice condensation occurs when the gas temperature $T$ becomes
$< T_{\rm ice}$ for the first time at individual $r$.
As the gas disk further expands, $T$ in the outer regions becomes well below $T_{\rm ice}$.
However, icy grains do not condense there, because the ices have already condensed and the gas there is free of water vapor.
The maximum radius $r_{\rm max}$ of the ice condensation is estimated 
by the intersection of Eq.~(\ref{eq:Sigma_ice}) and 
the envelope curve of superposition of $\Sigma_{\rm g}$-$r$ curves 
at different times (Figure 2a).
It is given by (Methods, Eq.~(\ref{eq:envelope}))
\begin{equation}
r_{\rm max} \simeq 20 \, 
 \left[\beta \left(\frac{\langle r_{\rm d,imp} \rangle}{2\, r_{\rm U}}\right)^{-5/4}
                              \left(\frac{M_{\rm d,imp}}{10^{-2} M_{\rm U}}\right)\right]^{1/4} r_{\rm U}.
\label{eq:rmax}
\end{equation}

From Eqs.~(\ref{eq:Sigma_ice}) and (\ref{eq:rmax}), the total condensed ice mass is
\begin{eqnarray}
M_{\rm ice} & \simeq & 
\int_{r_{\rm U}}^{r_{\rm max}} 2\pi r \Sigma_{\rm ice} d r
\simeq 0.58 \times 10^{-4} \, \beta^{1/8} \gamma_{03}
\left(\frac{\langle r_{\rm d,imp} \rangle}{2\, r_{\rm U}}\right)^{-5/4}
                              \left(\frac{M_{\rm d,imp}}{10^{-2} M_{\rm U}}\right)^{7/8}
\,M_{\rm U},
 \label{eq:M_ice}
\end{eqnarray}
which is consistent with the current total mass of Uranian satellites ($\simeq 1.0 \times 10^{-4}\,M_{\rm U}$).
Although the turbulent viscosity parameter $\alpha$ is uncertain, 
the $\alpha$-dependence of $M_{\rm ice}$ and $r_{\rm max}$ are very weak 
($\alpha \propto \beta)$.
Thus, we have demonstrated that
the compact ($\langle r_{\rm d,imp} \rangle \sim 2\,  r_{\rm U}$) 
and massive ($M_{\rm d,imp} \sim 10^{-2} M_{\rm U}$)
initial disk produces
the condensed ice confined at a distant place, $r_{\rm max}\sim 20 \,r_{\rm U}$
with the highly reduced total mass ($\sim 10^{-4} \,M_{\rm U}$).
This result clearly solves the problem of 
a too massive and too compact impact-generated disk.

Once (sub-micron) icy grains condense in the disk, they coagulate with each other. 
In general, as the icy particles grow, 
the particles drift inward by the aerodynamic gas drag \cite{Nakagawa86}.
However, the disk gas density is depleted so severely before
the ice condensation that the growth is much faster than the drift
(see Methods) and
km-sized ``satellitesimals" are formed in situ without radial drift.
Due to the disk gas depletion, ``type I migration" of proto-satellites caused by the torque from density waves 
in the disk would not be important, either (Methods).
Therefore, the satellitesimals and satellites are formed in situ.

The vaporization of rocks occurs at $T > 2000\,{\rm K}$ \cite{Melosh07}.
Owing to the high vaporization/condensation temperature, silicate (rock components) grains would quickly re-condense,
during the disk is still massive and compact.
Our model produces naturally an enhanced rock to ice ratio of the satellites 
because the ices condense after two orders of magnitude reduction
of water vapor, while the rocks condense before significant reduction.
Although the silicates condense only in inner region,
they would also spread uniformly in the disk.
Because silicate particles are not sticky at silicate-silicate collisions \cite{Blum00},
they do not grow beyond $\sim 100 \mu$m and radially spreads with the turbulent viscous dissipation in the disk, unless the turbulence is very weak (see Method).  
After the disk cools down and ice condensation starts, silicate particles can stick to icy particles
or ices condense to the silicate particle surface
beyond the ice line one after,
which could potentially account for a relatively uniform rock to ice ratio ($\sim O(1)$)  of all the satellites.
Thus, our model may also solve the small rock to ice ratio 
raised by previous simulations
\cite{Slattery92,Kegerreis18,Reinhardt19}, although
more detailed investigation is needed.

The condensed ice mass distribution peaks strongly at $\sim r_{\rm max}$.
This is consistent with the mass-orbit distribution of Uranian satellites (Fig.~1).
We have performed a direct 3D N-body simulation from 
10000 bodies with the individual masses $0.92 \times 10^{-8} M_{\rm U}$ 
that follow the ice distribution given by Eq.~(\ref{eq:Sigma_ice}) with $r_{\rm max}=20 \,r_{\rm U}$
and $\beta=\gamma_{03}=1$ (Methods).
Note that pebble accretion is negligible in our system (Methods).
The result reproduces the mass-orbit configuration of 
the current Uranian satellites in Fig. 1.
In a longer run, more consistent result would be obtained (see the caption).
Because orbital migration of satellites is not important, 
the satellites are not trapped in resonant orbits,
and the mass of accreted satellites is consistent with 
the isolation mass in oligarchic growth model \cite{Kokubo00}, given by 
(Methods, Eq.~(\ref{eq:miso}))
\begin{equation}
\frac{m_{\rm iso}}{M_{\rm U}} 
\simeq 0.74 \times 10^{-4} \beta^{-3/2} \gamma_{03}^{3/2}
\left(\frac{r}{20\,r_{\rm U}}\right)^{21/4}.
\label{eq:miso0}
\end{equation}
We also performed N-body simulations from
ordinary $\Sigma_{\rm ice}$-distributions with a negative radial gradient 
and robustly showed that 
a positive gradient of $\Sigma_{\rm ice}$ is required to reproduce 
the current mass-orbit configuration \cite{Ishizawa}.
   
We have shown that the current Uranian major satellites are 
beautifully reproduced by the derived analytical formulas 
based on viscous spreading and cooling
of the disk generated by an impact that is
constrained by the spin period and the tilted spin,
independent of details of the initial disk parameters.
Although we have focused on Uranus, the model here 
provides a general scenario for satellite formation around ice giants
with the scaling by the mass and the physical radius of a central planet,
which is totally different from satellite formation scenarios around terrestrial planets and gas giants.
It could also be applied for the inner region of Neptune's satellite system,
where we can neglect the effect of Triton that may have been captured \cite{Agnor06}.
Observations suggest that many of discovered super-Earths in exoplanetary systems may consist of abundant water ice, even in close-in (warm) orbits \cite{Rogers15}.
The model here may also give a lot of insights into possible icy satellites
of super-Earths.

\subsection{Acknowledgement}
This study was supported by 
MEXT “Exploratory Challenge on Post-K computer” (hp180183, hp190143),
“Priority Issue on post-K computer” (hp190156), 
JSPS KAKENHI 15H02065, 19K03950, and by MEXT KAKENHI 18H05438.
N-body simulation in this work was carried out at the Yukawa Institute Computer Facility.

\begin{methods}
\section*{Theory and numerical analysis.} 

\noindent
{\it Approximate self-similar solution to viscous diffusion equation.}

The analytical self-similar solution to Eq.~\ref{eq:surf_density_evol2} is given by
\cite{Lynden-Bell74, Hartmann98} 
\begin{eqnarray}
\Sigma_{\rm g} & \propto & t_*^{-\frac{5/2-\zeta}{2-\zeta}}
r^{-\zeta} \exp \left[ - \left(\frac{r}{r_{\rm d0}}\right)^{2-\zeta}t_*^{-1}\right].
\label{eq:self-similar}
\end{eqnarray}
where $\zeta = d\ln \nu /d \ln r$,
$t_* = 1+t/t_{\rm diff}$, 
\begin{equation}
t_{\rm diff} = \frac{1}{3(2-\zeta)^2} 
\left( \frac{r^2}{\nu} \right)_{r_{\rm d0}},
\label{eq:tdiff_zeta}
\end{equation}
and ``( )$_{r_{\rm d0}}$" means the value at $r_{\rm d0}$.
The surface density is $\propto r^{-\zeta}$ for $r \ll r_{\rm d}
= r_{\rm d0} \, t_*^{\;1/(2-\zeta)}$
and it exponentially decays for $r \ga r_{\rm d}$,
so that $r_{\rm d}$ is the characteristic disk radius.
In the case of our simple viscous heating model (Eq.~(\ref{eq:Tvis})), 
$\nu \sim \alpha c_s^2/ \Omega \propto T \, r^{3/2}
\propto \Sigma_{\rm g}^{1/3} \, r$.
In inner disk regions, the disk accretion is steady
and its rate is independent of $r$, that is, $\Sigma_{\rm g} \nu$ is independent of $r$.
In this case, $\nu \propto r^{3/4}$.
With $\zeta=3/4$, the self-similar solution given by Eq.~(\ref{eq:self-similar}) is
\begin{eqnarray}
\Sigma_{\rm g} & = & \Sigma_{\rm g,U0} \,t_*^{-7/5} 
\left(\frac{r}{r_{\rm U}}\right)^{\,-3/4}
\exp \left[ - \left(\frac{r}{r_{\rm d0}}\right)^{5/4} t_*^{-1}\right], 
\label{eq:Sigma_selfsim34}
\end{eqnarray}
where $r_{\rm U}$ is Uranian physical radius
given by $r_{\rm U} \simeq 2.5 \times 10^7$ m, 
and $\Sigma_{\rm g,U0}$ is the initial disk surface density at $r=r_{\rm U}$.

In the original self-similar solution, $t_{\rm diff}$ (Eq.~(\ref{eq:tdiff_zeta})) is a constant with time.
However, in our case, $\nu \propto \Sigma_{\rm g}^{1/3}$.
As the disk viscously expands and $\Sigma_{\rm g}$ decreases,
$\nu$ at $r=r_{\rm d0}$ in Eq.~(\ref{eq:tdiff_zeta}) also decreases.
As a result, $t_{\rm diff}$ increases.
Because we are concerned with $t > t_{\rm diff}$,
$t_* \propto t_{\rm diff}^{-1} \propto \nu \propto \Sigma_{\rm g}^{1/3}$.
Taking this effect into account, Eq.~(\ref{eq:Sigma_selfsim34}) suggests
$\Sigma_{\rm g} \propto t_{*0}^{-7/5} \Sigma_{\rm g}^{(-7/5) \times (1/3)}$,
that is, $\Sigma_{\rm g} \propto t_{*0}^{-21/22}$,
where $t_{*0} = 1+t/t_{\rm diff0}$, and $t_{\rm diff0}$ is defined by
quantities at $t=0$ as (Eq.~(\ref{eq:tdiff_zeta}) with $\zeta = 3/4$)
\begin{equation}
t_{\rm diff0} = \frac{16}{75} \left( \frac{r^2}{\nu}\right)_{r_{\rm d0},t=0}.
\label{eq:t_diff0}
\end{equation}
%The numerical factor is obtained by Eq.~(\ref{eq:tdiff_zeta}) with $\zeta = 3/4$.
Because $\Sigma_{\rm g} \propto t_{*0}^{-21/22}$ and 
$t_* \propto t_{*0} \Sigma_{\rm g}^{1/3} \propto t_{*0}^{15/22}$,
the final formula is 
\begin{eqnarray}
\Sigma_{\rm g} & = & \Sigma_{\rm g,U0} \,t_{*0}^{-21/22} \left(\frac{r}{r_{\rm U}}\right)^{\,-3/4}
\exp \left[ - \left(\frac{r}{r_{\rm d0}} \right)^{5/4}t_{*0}^{-15/22} \right], 
\label{eq:Sigma_selfsim3} \\
t_{*0} & = & 1+t/t_{\rm diff0}.
\label{eq:t_star0}
\end{eqnarray}
Although this formula is no longer a strict self-similar solution,
it reproduces the numerical solution well, as shown in Fig.~2 in the main text.

\vspace{0.7cm}
\noindent
{\it Initial relaxation to the self-similar solution.}

The impact-generated disk is quickly relaxed to the analytical quasi-steady self-similar solution 
(Eq.~\ref{eq:Sigma_selfsim3}). 
The parameters $r_{\rm d0}$ and $\Sigma_{\rm g,U0}$ 
in the self-similar solution are estimated by the total mass ($M_{\rm d,imp}$) and 
the angular momentum ($J_{\rm d,imp}$) of the impact-generated disk.
In general, SPH simulations show that the impact-generated disk
is compact and 
the mean radius is $\langle r_{\rm d,imp} \rangle \sim 2 r_{\rm U}$ \cite{Slattery92, Kegerreis18,Reinhardt19},
where $\langle r_{\rm d,imp} \rangle $ is defined with the specific angular momentum,
$j_{\rm d,imp}=J_{\rm d,imp}/M_{\rm d,imp}$, by $\langle r_{\rm d,imp} \rangle  = (j_{\rm d,imp}/r_{\rm U}^2 \Omega_{\rm U})^2 r_{\rm U}$.
The value of $\langle r_{\rm d,imp} \rangle$ is larger for a less steep disk surface density distribution. 
In the SPH impact simulations, debris particles generally have eccentric orbits.
Since the orbits should be eventually circularized, conserving 
angular momentum, 
we define $\langle r_{\rm d,imp} \rangle$ with the assumption that the orbits are circular,
while $j_{\rm d,imp}$ must be calculated from debris particles in eccentric orbits
in the simulation results. 

Because the radial gradient of the disk surface density is generally very steep,
the disk expands to a self-similar distribution, 
almost keeping the total disk angular momentum.
While the total angular momentum is conserved,
the innermost disk generally tends to spiral in by losing angular momentum.
The one-dimensional diffusion simulations in this paper 
show that a half of the mass inside $\langle r_{\rm d,imp} \rangle$
falls onto the planet until the disk
settles down to the self-similar solution.
If we consider the disk surface density distribution just after the impact as
$\Sigma_{\rm g} \propto r^{-3}$ with a truncation at $r = 10 \, r_{\rm U}$,
which is suggested by SPH simulations,
the initial mass of the impact-generated disk ($M_{\rm d,imp}$) is decreased by $\sim 20\%$
in the early relaxation.
Using $J_{\rm d,imp}$ of the impact-generated disk and 
the modified disk mass $0.8 M_{\rm d,imp}$, we can evaluate  
$r_{\rm d0}$ and $\Sigma_{\rm g,U0}$ in 
the self-similar solution as follows.

The total disk mass and angular momentum of the self-similar solution are
\begin{eqnarray} 
M_{\rm d,ss} & = & \int^{\infty}_{r_{\rm U}} 2 \pi r \Sigma_{\rm g} dr =  
\frac{8 \pi}{5} r_{\rm U}^2 \Sigma_{\rm g,U0}  \, \left(\frac{r_{\rm d0}}{r_{\rm U}}\right)^{\,5/4} 
e^{-(r_{\rm d0}/r_{\rm U})^{-5/4}} \nonumber\\
 & \simeq & \frac{8 \pi}{5} r_{\rm U}^2 \Sigma_{\rm g,U0}  \, \left(\frac{r_{\rm d0}}{r_{\rm U}}\right)^{\,5/4} \times 0.776, 
\label{eq:Mg}
\\ 
 J_{\rm d,ss} & = & \int^{\infty}_{r_{\rm U}} 2 \pi r \Sigma_{\rm g} \sqrt{G M_{\rm U} r} \, dr  = \frac{8 \pi}{5} r_{\rm U}^4 \Omega_{\rm U} \Sigma_{\rm g,U0} \, \left(\frac{r_{\rm d0}}{r_{\rm U}}\right)^{\,7/4} \,
   \Gamma \left(\frac{7}{5},\left(\frac{r_{\rm d0}}{r_{\rm U}}\right)^{-5/4}\right)
   \nonumber\\ 
    & \simeq & \frac{8 \pi}{5} r_{\rm U}^4 \Sigma_{\rm g,U0} \Omega_{\rm U} \, \left(\frac{r_{\rm d0}}{r_{\rm U}}\right)^{\,7/4} \, \times 0.797,
\label{eq:Jg}
 \end{eqnarray}
where $\Gamma$ is a 2nd-kind incomplete gamma function,
$\Omega_{\rm U}$ is the disk orbital frequency at $r=r_{\rm U}$,
and we used $r_{\rm d0}/r_{\rm U} \sim 3$ to evaluate
$e^{-(r_{\rm d0}/r_{\rm U})^{-5/4}}$ and $ \Gamma \left(\frac{7}{5},(r_{\rm d0}/r_{\rm U})^{-5/4}\right)$.
From Eqs.~(\ref{eq:Mg}) and (\ref{eq:Jg}), the mean specific angular momentum
of the self-similar solution is given by
\begin{eqnarray}
j_{\rm d,ss} & \simeq & \frac{J_{\rm d,ss}}{M_{\rm d,ss}} = 1.03 \, \left(\frac{r_{\rm d0}}{r_{\rm U}}\right)^{1/2} \Omega_{\rm U} r_{\rm U}^2. 
\label{eq:jdisk}
\end{eqnarray}
Because $j_{\rm d,ss}=J_{\rm d,ss}/M_{\rm d,ss} \simeq J_{\rm d,imp}/0.8 M_{\rm d,imp} \simeq 1.25 \, j_{\rm d,imp}$,
\begin{eqnarray}
r_{\rm d0} & \simeq & 1.47 \left(\frac{j_{\rm d,imp}}{r_{\rm U}^2 \Omega_{\rm U}}\right)^2 r_{\rm U}
= 1.47 \, \langle r_{\rm d,imp} \rangle.
\label{eq:rd0_jdisk}
\end{eqnarray}
From Eq.~(\ref{eq:Mg}) with $M_{\rm d,ss} \sim 0.8 \, M_{\rm d,imp}$, 
the surface density of the self-similar solution after the initial relaxation of
the impact-generated disk is 
\begin{eqnarray}
\Sigma_{\rm g,U0} & \simeq & 0.256 \left(\frac{r_{\rm d0}}{r_{\rm U}}\right)^{-5/4}
                              \left(\frac{M_{\rm d,ss}}{r_{\rm U}^2}\right) \nonumber \\
%                              \simeq & 0.126\left(\frac{j_{\rm disk}}{r_{\rm U}^2 \Omega_{\rm U}}\right)^{-5/2}                              \left(\frac{M_{\rm disk}}{r_{\rm U}^2}\right) 
                          & \simeq & 6.5 \times 10^7 \left(\frac{\langle r_{\rm d,imp} \rangle}{2\, r_{\rm U}}\right)^{-5/4}
                              \left(\frac{M_{\rm d,imp}}{10^{-2} M_{\rm U}}\right) \,\rm kg/m^2.
                              \label{eq:Sigmag0_jdisk}
\end{eqnarray}
In the case of the impact-generated disk with 
$\Sigma_{\rm g} = \Sigma_{\rm g,imp0} (r/r_{\rm U})^{-3}$ with a truncation at $r = 10 \, r_{\rm U}$,
$M_{\rm d,imp} = 0.9 \times 2\pi \Sigma_{\rm g,imp0} \, r_{\rm U}^2$
and $\langle r_{\rm d,imp} \rangle \simeq 2.25 \, r_{\rm U}$,
so that $r_{\rm d0} \simeq 3.3\,  r_{\rm U}$ and $\Sigma_{\rm g,U0} \simeq 0.26 \,\Sigma_{\rm g,imp0}$.

As discussed in the main text,
to evaluate the outer limit of the ice condensation, 
the envelope curve of superposition of $\Sigma_{\rm g}$-$r$ curves at all the different times
is important.
The $\Sigma_{g}$-distribution of the analytical solution starts exponentially declining
at $r_{\rm d} \sim r_{\rm d0} \, t_{*0}^{12/22}$ and 
the absolute values of $\Sigma_{\rm g}$ at the same $r$ scale by $t_{*0}^{-21/22}$,
while $\Sigma_{\rm g}$ further decreases in proportion to $r_{\rm d}^{-3/4} \propto t_{*0}^{-9/22}$, 
as shown in Eq.~(\ref{eq:Sigma_selfsim3}).
Therefore, the envelope curve is given by
\begin{equation}
\Sigma_{\rm g,env} 
\simeq \Sigma_{\rm g,U0} \left(\frac{r}{r_{\rm U}}\right)^{-[(21+9)/22]/(12/22)}
\simeq 
6.5 \times 10^7 \left(\frac{\langle r_{\rm d,imp} \rangle}{2\, r_{\rm U}}\right)^{-5/4}
                              \left(\frac{M_{\rm d,imp}}{10^{-2} M_{\rm U}}\right)
%3 \times 10^7\, \left( \frac{\Sigma_{\rm g,U0}}{3 \times 10^7 \, {\rm kg/m^2}}\right)
\left(\frac{r}{r_{\rm U}}\right)^{-5/2} \;{\rm kg/m^2}.
\label{eq:envelope}
\end{equation}
It agrees with the numerical result in Fig.~2.
The intersection radius between $\Sigma_{\rm g,env}$ and
$\Sigma_{\rm g}$ at the ice condensation (Eq.~(\ref{eq:Sig_T})) is given by
\begin{equation}
% r_{\rm intsec} \simeq 20 \, 
r_{\rm max} \simeq 20 \, 
 \left[\beta \left(\frac{\langle r_{\rm d,imp} \rangle}{2\, r_{\rm U}}\right)^{-5/4}
                              \left(\frac{M_{\rm d,imp}}{10^{-2} M_{\rm U}}\right)\right]^{1/4} r_{\rm U}.
\label{eq:rmax2}
\end{equation}

\vspace{0.7cm}
\noindent
{\it Icy grain growth/drift and disk diffusion timescales.}

Here we show that the growth of condensed icy particles is much faster than their radial drift
and the gas disk diffusion.
Thereby, the condensed icy grains quickly grow in situ to km-sized ``satellitesimals," 
which are building blocks of satellites, in the H/He gas disk.
We estimate the timescales of individual processes at $r \sim 20 \, r_{\rm U}$  
because most of the icy grains condense there.

{\it Disk diffusion timescale:}

We consider a disk with a characteristic radius of $r_{\rm d0}$ and
a turbulent viscosity of $\alpha c_s^2 \Omega^{-1}$,
where $c_s$ is the local sound velocity of the disk gas,
$\Omega$ is the local orbital frequency of the gas, and $\alpha$ is
a parameter to represent the strength of turbulence ($\alpha \ll 1$) \cite{Shakura73}.  
From Eqs.~(\ref{eq:Sigma_selfsim3}) and (\ref{eq:t_star0}),
the disk diffusion timescale is given by
\begin{equation}
t_{\rm diff} \sim \frac{\Sigma_{\rm g}}{d \Sigma_{\rm g}/dt}
\simeq t_{\rm diff0}\, t_{*0} \simeq \max(t_{\rm diff0}, t),
\label{eq:t_diff00}
\end{equation}
where $t_{\rm diff0}$ is the initial disk diffusion timescale given by
\begin{equation}
t_{\rm diff0} \sim \left( \frac{16\, r^2}{75\,\nu}\right)_{r_{\rm d0},t=0} 
\sim \frac{16}{75\, \alpha} 
\left[ \left(\frac{c_s}{v_{\rm K}}\right)^{-2} \Omega^{-1}\right]_{r_{\rm d0},t=0}.
\label{eq:t_diff0}
\end{equation}
The value of $c_s/v_{\rm K}$, which is equivalent to the disk aspect ratio, is
\begin{equation}
\frac{c_s}{v_{\rm K}} \simeq 0.0564 
\left(\frac{T}{240 \rm K}\right)^{1/2}\left(\frac{r}{r_{\rm U}}\right)^{1/2},
\label{eq:cs_vk}
\end{equation}
where we use the mean molecular weight $\sim 2.8$.
Substituting Eqs.~(\ref{eq:Sig_g0}) and (\ref{eq:T_Sigma}) into Eq.~(\ref{eq:cs_vk}), for 
the initial self-similar disk after the relaxation, 
\begin{equation}
\left( \frac{c_s}{v_{\rm K}} \right)_{r_{\rm d0},t=0} \simeq 0.416
\left(\frac{\langle r_{\rm d,imp} \rangle}{2\, r_{\rm U}}\right)^{-5/8}
\left(\frac{M_{\rm d,imp}}{10^{-2} M_{\rm U}}\right)^{1/2}
\left(\frac{r_{\rm d0}}{r_{\rm U}}\right)^{1/8}.
\label{eq:cs_vk0}
\end{equation}
Adopting a typical impact-generated disk
with $\langle r_{\rm d,imp} \rangle \sim 2\, r_{\rm U}$ and $M_{\rm d,imp} \sim 10^{-2} M_{\rm U}$
and the corresponding relaxed disk with 
$r_{\rm d0}\sim 3 \, r_{\rm U}$, and scaling $\Omega^{-1}$ at $r \sim 20 r_{\rm U}$,
Eq.~(\ref{eq:t_diff0}) reads as
\begin{equation}
t_{\rm diff0} \sim
54 \left(\frac{\alpha}{10^{-3}} \right)^{-1}
\Omega^{-1}.
\label{eq:t_diff02}
\end{equation}

Because $\Sigma_{\rm g} \propto t_{\rm *0}^{-21/22}$,
the time from the initial $\Sigma_{\rm g}$ given by Eq.~(\ref{eq:Sig_g0}) 
to $\Sigma_{\rm g}$ at the ice condensation given by Eq.~(\ref{eq:Sigma_ice})
at $r \sim 20\, {\rm au}$ is
\begin{equation}
t \simeq t_{\rm *0} \, t_{\rm diff0} \simeq
\left(
\frac{\Sigma_{\rm g, Eq.(\ref{eq:Sig_g0})}}{\Sigma_{\rm g, Eq.(\ref{eq:Sigma_ice})}}\right)^{22/21} t_{\rm diff0} 
 \simeq 1.7 \times 10^4 
\left[\beta \left(\frac{\langle r_{\rm d,imp} \rangle}{2\, r_{\rm U}}\right)^{-5/4}
\left(\frac{M_{\rm d,imp}}{10^{-2} M_{\rm U}}\right)\right]^{22/21} t_{\rm diff0}.
\end{equation}
Therefore, the disk diffusion timescale at the ice condensation is
\begin{equation}
t_{\rm diff} \simeq t \simeq 9.2 \times 10^5 
\left[\beta \left(\frac{\langle r_{\rm d,imp} \rangle}{2\, r_{\rm U}}\right)^{-5/4}
\left(\frac{M_{\rm d,imp}}{10^{-2} M_{\rm U}}\right)\right]^{22/21}
\left(\frac{\alpha}{10^{-3}} \right)^{-1} \Omega^{-1}.
\label{eq:t_diff03}
\end{equation}

{\it Drift timescale of icy particles due to gas drag:}

The condensed icy grains coagulate with each other. 
As the icy particles grow, their motions become less coupled to the disk gas.
The degree of the decoupling is represented by Stokes number,
${\rm St} = t_{\rm stop}\Omega$, where $t_{\rm stop}$ is the stopping time due to 
aerodynamic gas drag.
The disk gas rotates slower than the particles 
by a small fraction of $\eta \sim (c_s/v_{\rm K})^2$ $(\ll 1)$.
By the drag from the slower rotating disk gas, 
the particles drift inward with the drift timescale given by \cite{Nakagawa86}:
\begin{equation}
t_{\rm drift} \simeq \frac{r}{v_r}
\simeq \frac{r}{2 \eta \, v_{\rm K}}\frac{1+\st^2}{\st}
\simeq 0.5 \left(\frac{c_s}{v_{\rm K}}\right)^{-2} \frac{1+\st^2}{\st} \Omega^{-1},
\label{eq:t_drift}
\end{equation}
where $v_r$ is the radial drift velocity.
At $r \sim 20 \, r_{\rm U}$, 
$(c_s/v_{\rm K})^{-2} \sim 16$ (Eq.~(\ref{eq:cs_vk})). 
The drift is the fastest at $\st \sim 1$.

{\it Growth timescale of icy particles:}

The growth times scale (mass-doubling timescale) 
of icy particles with ${\rm St} \la 1$ is given by
\begin{equation}
t_{\rm grow} \sim \frac{1}{n \pi R^2 \Delta v},
\label{eq:t_grow}
\end{equation}
where $R$ is the particle physical radius, $n$ is their spatial number density, 
\begin{eqnarray}
n & = & \frac{\rho_{\rm p}}{(4\pi/3)\rho_{\rm mat} R^3},  \label{eq:n}
\end{eqnarray}
$\rho_{\rm p}$ and $\rho_{\rm mat}$ are the spatial and material
densities of the particles, and $\Delta v$ is the relative velocity between the particles 
\cite{OrmelCuzzi07},
\begin{eqnarray}
\Delta v
\simeq \left(3\alpha \, \st \right)^{1/2} c_s. \label{eq:vcol}
\end{eqnarray}
The icy particle spatial density is given by their surface density $\Sigma_{\rm ice}$ as
\cite{Dubrulle95}
\begin{equation} 
\rho_{\rm p} \simeq \frac{\Sigma_{\rm ice}}{\sqrt{2\pi} h_{\rm p}} 
\simeq \frac{\Sigma_{\rm ice}}{\sqrt{2\pi} h_{\rm g}} \left(1+\frac{\rm St}{\alpha}\right)^{1/2},
\label{eq:rho_p}
\end{equation}
where $h_{\rm p}$ and $h_{\rm g}$ are the particle and the gas vertical scale heights.
Substituting Eqs.~(\ref{eq:n}), (\ref{eq:vcol}), and (\ref{eq:rho_p}) into Eq.~(\ref{eq:t_grow}),
we obtain
\begin{equation}
t_{\rm grow} \sim \frac{4\sqrt{2\pi}}{3\sqrt{3}}
\frac{\rho_{\rm mat}R}{\sqrt{\st(\st +\alpha)}\, \Sigma_{\rm ice}}\Omega^{-1},
\label{eq:t_grow2}
\end{equation}
where we used the disk gas scale height is given by $h_{\rm g} \sim c_s \Omega^{-1}$.

In the situation we are considering,
the drag law is mostly in Stokes drag regime.
In this case, the Stokes number is given by
\begin{equation}
{\rm St} \sim \frac{4 \rho_{\rm mat} \sigma_{\rm coll} R^2 \Omega}{9 \mu_{\rm HHe} \, m_{\rm H} \, c_s}
\sim  1.5 \times 10^{-6} \left( \frac{T_{\rm ice}}{240\,{\rm K}} \right)^{-1/2}\left(\frac{R}{\mu {\rm m}}\right)^2 \left(\frac{r}{r_{\rm U}}\right)^{\, -3/2}, 
\label{eq:St}
\end{equation}
where we used $\rho_{\rm mat} \sim 10^3 \; {\rm kg/m^3}$,
$\mu_{\rm HHe} \sim 2.4$ is the mean molecular weight for H-He gas, 
$m_{\rm H}\sim 1.67 \times 10^{-21}{\rm kg}$ is the hydrogen mass, and 
$\sigma_{\rm col}\sim 2 \times 10^{-11} {\rm m^2}$ is the collision cross section.
Substituting Eqs.~(\ref{eq:St}) and (\ref{eq:Sigma_ice})
into Eq.~(\ref{eq:t_grow2}), we obtain
\begin{equation}
t_{\rm grow} \sim 1
\left(\frac{\st + \alpha}{10^{-4}}\right)^{-1/2}
\left(\frac{\gamma}{0.3}\right)^{-1} 
\left(\frac{\alpha}{10^{-3}}\right)
\left( \frac{T_{\rm ice}}{240\,{\rm K}} \right)^{-11/4}
\left(\frac{r}{r_{\rm U}}\right)^{\,-3/4}
\Omega^{-1}.
\label{eq:t_grow3}
\end{equation}

{\it Timescale comparison:}

Because $c_s < v_{\rm K}$ and $\alpha \ll 1$, 
\begin{equation}
t_{\rm grow} \ll t_{\rm drift}, t_{\rm diff}.
\label{eq:timescales}
\end{equation}
Around $\st \sim 1$,
\begin{equation}
t_{\rm grow} \ll t_{\rm drift} \ll t_{\rm diff}.
\label{eq:timescales2}
\end{equation}
These results imply that 
the condensed icy grains quickly grow to km-sized satellitesimals 
in situ in the H/He gas disk.
The satellitesimal motions are decoupled from the disk gas.

\vspace{0.7cm}
\noindent
{\it Ice condensation.}

Icy grains condense when the vapor pressure exceeds 
the vapor saturation pressure.
Because the vapor saturation pressure depends sensitively on temperature,
the condensation condition is often described by $T < T_{\rm ice}$,
where $T_{\rm ice}$ is the condensation temperature given by 
\cite{Lichtenegger91}
\begin{equation} 
T_{\rm ice} \simeq \frac{A}{B - \log_{10} (P_{\rm H2O} {\rm [Pa]})} \; [\rm K]
\label{eq:snowline}
\end{equation}
with 
\begin{equation} 
A \simeq 2633 \; ; \; B \simeq 12.06,
\end{equation}
where $P_{\rm H2O}$ is the partial pressure of water vapor in the disk, 
given by
\begin{equation}
P_{\rm H2O} = \gamma \, \frac{\mu_{\rm all}}{\mu_{\rm H2O}} P \simeq 0.156 \, \gamma \, P,
\label{eq:PH2O}
\end{equation}
where $P$ is the total pressure, 
$\gamma = \Sigma_{\rm H2O}/\Sigma_{\rm g}$,
and $\mu_{\rm all}\simeq 2.8$ and $\mu_{\rm H2O} = 18$ are
the total and H$_2$O mean molecular weight.

The total pressure is
\begin{eqnarray}
P & = & \rho_{\rm g} c_s^2 = \frac{\Sigma_{\rm g}}{\sqrt{2\pi}} c_s \Omega
 \simeq 61.9 \left(\frac{\alpha}{10^{-3}}\right)^{-1} 
 \left( \frac{T}{240\,{\rm K}} \right)^{7/2} \; \; {\rm Pa}, 
 \label{eq:PT} 
\label{eq:rho_gA}
\end{eqnarray}
where we used 
\begin{equation}
c_s \simeq 8.41 \times 10^2\,(\mu_{\rm all}/2.8)^{-1/2}(\rm T/240K)^{1/2} \;  {\rm m/s},
\label{eq:cs}
\end{equation}
and $\Sigma_{\rm g}$ obtained by Eq.~(\ref{eq:Tvis}),
\begin{equation}
\Sigma_{\rm g} \simeq 4.02 \times 10^{2} \left(\frac{\alpha}{10^{-3}}\right)^{-1} 
\left( \frac{T}{240\,{\rm K}} \right)^3 \left(\frac{r}{r_{\rm U}}\right)^{3/2} \; {\rm kg/m^2},
\label{eq:Sig_T}
\end{equation}
Thereby,
\begin{equation} 
P_{\rm H2O} = 0.156 \, \gamma P 
\simeq 9.66 \gamma \left(\frac{\alpha}{10^{-3}}\right)^{-1} 
 \left( \frac{T}{240\,{\rm K}} \right)^{7/2} \;\; {\rm Pa}, 
 \label{eq:PT0}
\end{equation}
From Eqs.~(\ref{eq:snowline}) and (\ref{eq:PT0}) with $T=T_{\rm ice}$,
we found 
\begin{eqnarray} 
T_{\rm ice} & \simeq & 
\frac{2633}{12.06 -  0.98 - \log_{10} \left[ \frac{\gamma}{0.3} \left(\frac{\alpha}{10^{-3}}\right)^{-1}  \right] } \; \;\rm K \nonumber\\
 & \simeq & \frac{238}{1 -\frac{1}{11.08} \log_{10} \left[ \frac{\gamma}{0.3} \left(\frac{\alpha}{10^{-3}}\right)^{-1}  \right] }\;\; {\rm K}
\simeq  238 +  21 \log_{10} \left[ \frac{\gamma}{0.3} \left(\frac{\alpha}{10^{-3}}\right)^{-1}  \right] \; \; {\rm K}.
\label{eq:T_snow} 
\end{eqnarray}
Note that the $r$-dependence vanishes for $T_{\rm ice}$ in our disk model.

% \vspace{0.7cm} \noindent {\it Silicates condensation.} {\bf In the similar way to the ice condensation temperature formula, the condensation temperature of silicate is approximated by \cite{Melosh07} \begin{equation} A \simeq 2 \times 10^4 \; ; \; B \simeq 11.5. \end{equation} Considering silicate vapor pressure, \begin{equation} T_{\rm rock} \simeq 2500 + 294 \log_{10} \left[ \frac{\gamma'}{0.003} \left( \frac{\alpha}{10^{-3}} \right)^{-1} \right] \; {\rm [K]}, \label{eq:T_snow_sil} \end{equation} where $\gamma'$ is the mass fraction of silicate vapor in the gas, which may be $\ll 1$.}

\vspace{0.7cm}
\noindent
{\it Barriers for silicate particle sticking.}

When collision velocity exceeds a threshold value ($\sim 1$ m/s),
silicate-silicate collisional sticking is inhibited by rebounding or fragmentation \cite{Blum00}. 
In the parameter range we consider, the particle collision velocity
induced by turbulence is given by Eqs.~(\ref{eq:vcol}) and (\ref{eq:cs}).
The maximum Stokes number of the particles that allows
the sticking is given by $v_{\rm bf} \sim \Delta v$ as
\begin{equation}
{\rm St}_{\rm max} \sim 
\frac{1}{3 \alpha} \left(\frac{v_{\rm bf}}{c_s}\right)^2
\sim 5 \times 10^{-4} \left(\frac{\alpha}{10^{-3}}\right)^{-1}
\left(\frac{v_{\rm bf}}{1 \,\rm m/s}\right)^2
\left(\frac{\mu_{\rm all}}{2.8}\right)
\left(\frac{T}{240 \rm K}\right)^{-1}.
\label{eq:St}
\end{equation}
Thus, silicates can grow only up to ${\rm St} \sim 5 \times 10^{-4}$ until 
$T$ deceases to ice condensation temperature $\sim 240$ K.
In the Stokes drag regime, it corresponds to the particle size of $\sim 100 \, \mu$m.
The silicate particles can form satellitesimals only after ices condense and
they stick to the icy particles or ices condense to their surface. 

\vspace{0.7cm}
\noindent
{\it N-body simulation:}

We perform 3D N-body simulation
from 10000 bodies (satellitesimals) with the individual masses $0.92 \times 10^{-8} M_{\rm U}$ with the predicted ice distribution given 
by Eq.~(\ref{eq:Sigma_ice}) with $r_{\rm max}=20 r_{\rm U}$
and $\beta=\gamma_{03}=1$.
Gravitational interactions of all the bodies are included.
Aerodynamical gas drag to satellitesimals and
type I migration due to disk-planet interactions is neglected as below.
Tidal interactions with Uranus are also neglected, because 
the timescale of our run is too short for the effect to be important.
We assume perfect accretion and the physical radii 
are increased by a factor of 2 to accelerate the growth. 
Small eccentricities and inclinations are given initially.
They are quickly relaxed by gravitational stirring and collision damping.  
Note that since there is no large reservoir of icy particles in outer region
of the disk and no icy particle supply from outside of the Uranian system,
pebble accretion is not effective and satellitesimals grow through
mutual collisions.   

When a proto-satellite grows, type I migration due to the torque
from the density waves in the gas disk can become important.
However, we show that its timescale is longer than disk diffusion timescale
and its effect is negligible.
The migration timescale of a satellite with mass $m$ is \cite{Tanaka02}
\begin{equation}
t_{\rm mig} \sim \frac{1}{2.7 + 1.1\times (3/4)}
\left(\frac{M_{\rm U}}{m}\right)
\left(\frac{M_{\rm U}}{\Sigma_{\rm g} r^2} \right)
\left(\frac{c_s}{v_{\rm K}}\right)^{2} \Omega^{-1}.
\label{eq:mig}
\end{equation}
Because type I migration is cased by a residual between the inner and outer disk torques and between Lindblad and corotation torques, the numerical factor depends on
the gas disk structure (sometimes it changes the sign).
However, the absolute value of the timescale is generally of the same order for any disk structure. 
At the ice condensation with $T \sim 240 \, \rm K$ at $r \sim \, 20 r_{\rm U}$, 
$c_s/v_{\rm K} \sim 0.25$ (Eq.~(\ref{eq:cs_vk})).
For $m/M_{\rm U}\sim 3\times 10^{-5}$ and
$\Sigma_{\rm g}r^2/M_{\rm U}\sim 10^{-4}$,
where we consider the most massive satellites,
%and used the surface density of H/He gas as $(1-\gamma)\Sigma_{\rm g}$ at the ice condensation and at $r=20 \,r_{\rm U}$ (Eq.~(\ref{eq:Sig_T})),
the type I migration timescale is
$t_{\rm mig} \sim 0.6 \times 10^7 \,\Omega^{-1}$.
Because $t_{\rm diff}$ at the ice condensation is $\sim 0.9 \times 10^6 \,\Omega^{-1}$
(Eq.~(\ref{eq:t_diff02}))
and the H/He gas should decay more when the large enough 
satellites grow from satellitesimals,
it is predicted that $t_{\rm mig} \gg t_{\rm diff}$.
Because $t_{\rm diff} \propto t_{*0} \propto \Sigma_{\rm g}^{-22/21}$
and $t_{\rm mig} \propto \Sigma_{\rm g}$, the relation of
$t_{\rm mig} \gg t_{\rm diff}$ does not change afterward.
% Furthermore, since $\Sigma_{\rm g} r^2 \sim m$, the disk gas may be modulated rather than the angular momentum of the proto-satellites is removed.   
Therefore, type I migration of proto-satellites is negligible.

\vspace{0.7cm}
\noindent
{\it Isolation mass in oligarchic growth.}

In the context of planet accretion, if orbital migration is neglected,
the planetary accretion is terminated when small bodies in the feeding zone
of the planet is consumed, and the planetary mass at that point is called ``isolation mass." \cite{Kokubo00}
In the system we consider here, the isolation mass ($m_{\rm iso}$) is defined by
\begin{equation}
m_{\rm iso} = 2 \pi r \Delta r \Sigma_{\rm ice},
\end{equation}
where $\Delta r$ is orbital distance between proto-satellites
and $\Delta r \sim 10 (2m_{\rm iso}/3M_{\rm U})^{1/3} r$. 
It is rewritten as
\begin{eqnarray}
\frac{m_{\rm iso}}{M_{\rm U}} 
\simeq \frac{10\times 2^{1/3}}{3^{1/3}} \left(\frac{2 \pi \Sigma_{\rm ice} r^2}{M_{\rm U}}\right)^{3/2} 
\simeq 0.74 \times 10^{-4} \beta^{-3/2} \gamma_{03}^{3/2}
\left(\frac{r}{20\,r_{\rm U}}\right)^{21/4}.
\label{eq:miso}
\end{eqnarray}
The steep radial gradient of $m_{\rm iso}$ explains 
the orbital configuration of the current Uranian satellites (Fig.~1).

\end{methods}

%\bibliography{uranus}

%% Here is the endmatter stuff: Supplementary Info, etc.
%% Use \item's to separate, default label is "Acknowledgements"

\end{narrow}

\end{document}